\documentclass{PoS}
\usepackage{verbatim}
\usepackage{amsmath}

\def\re {{\rm Re}}
\def\im {{\rm Im}}
\def\fns {\footnotesize}
\title{Measurement of the CKM angle $\gamma$ ($\phi_3$) using $B\to DK$ decays}

\ShortTitle{Measurement of the CKM angle $\gamma$ using $B\to D K$ decays}

\author{\speaker{Matteo Rama}\\
        INFN, Laboratori Nazionali di Frascati, Via E. Fermi 40, I-00044 Frascati, Italy\\
        E-mail: \email{matteo.rama@lnf.infn.it}}

\abstract{We present the status of the measurement of the CKM $CP$-violating phase $\gamma$~$(\phi_3)$ performed using $B\to D^{(*)}K^{(*)}$ decays. We review and compare the main experimental methods.}

\FullConference{Flavor Physics and CP Violation 2009,\\
		May 27 - June 1, 2009,\\
		Lake Placid, NY, USA}

\begin{document}

\section{Introduction}
\begin{figure}[h!]
\begin{center}
\includegraphics[width=7cm]{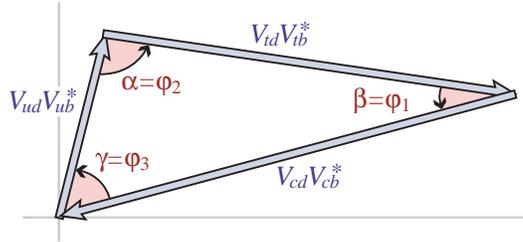}
\end{center}
\caption{Graphical representation of the unitarity constraint $V_{ud}V^*_{ub}+V_{cd}V^*_{cb}+V_{td}V^*_{tb}=0$ as a triangle in the complex plane~\cite{pdg2008}.}
\label{fig:UT}
\end{figure}

$CP$ violation was first established in $K^0_L\to\pi^+\pi^-$ decays in 1964~\cite{cpviol} and since then it has been observed in several $K$ and $B$ meson weak decays~\cite{pdg2008}. In the standard model of particle physics (SM) $CP$ violation in the quark sector of the weak interactions arises from a single irreducible phase in the Cabibbo-Kobayashi-Maskawa (CKM) matrix~\cite{ckm} that describes the mixing of the quarks. 
The unitarity of the CKM matrix imposes a set of relations among its elements, including the condition $V_{ud}V^*_{ub}+V_{cd}V^*_{cb}+V_{td}V^*_{tb}=0$ that defines a {\it unitarity triangle} in the complex plane, shown in Fig.~\ref{fig:UT}. Many measurements can be conveniently displayed and compared as constraints on sides and angles of this triangle. 
$CP$ violation is proportional to the area of the unitarity triangle and therefore it requires that all sides and angles be different from zero. 
The angle $\gamma\equiv\phi_3\equiv\text{arg}(-V_{ud}V_{ub}^*/V_{cd}V_{cb}^*)$ is at present the most difficult to measure.

An important goal of flavor physics is to over-constrain the CKM matrix.
The reason is twofold. First, it is desirable to determine its elements as precisely as possible because their values are fundamental parameters of the SM. Second, new physics (NP) effects could manifest themselves as inconsistencies between two or more measurements of the CKM parameters~\cite{cdr_superb}. The angle $\gamma$ can be measured in decays mediated by tree amplitudes, such as $B\to D^0K$: assuming that NP does not change the tree-level processes, its determination is not affected by effects beyond the SM\footnote{NP may appear in $D^0-\bar D^0$ mixing but the effect is expected to be small and can be taken into account~\cite{arxiv_0802_3201}.} and together with the measurement of $|V_{ub}/V_{cb}|$ provides a constraint 
that can be compared with those potentially sensitive to NP.

\begin{figure}[!h]
\begin{center}
\includegraphics[width=5cm]{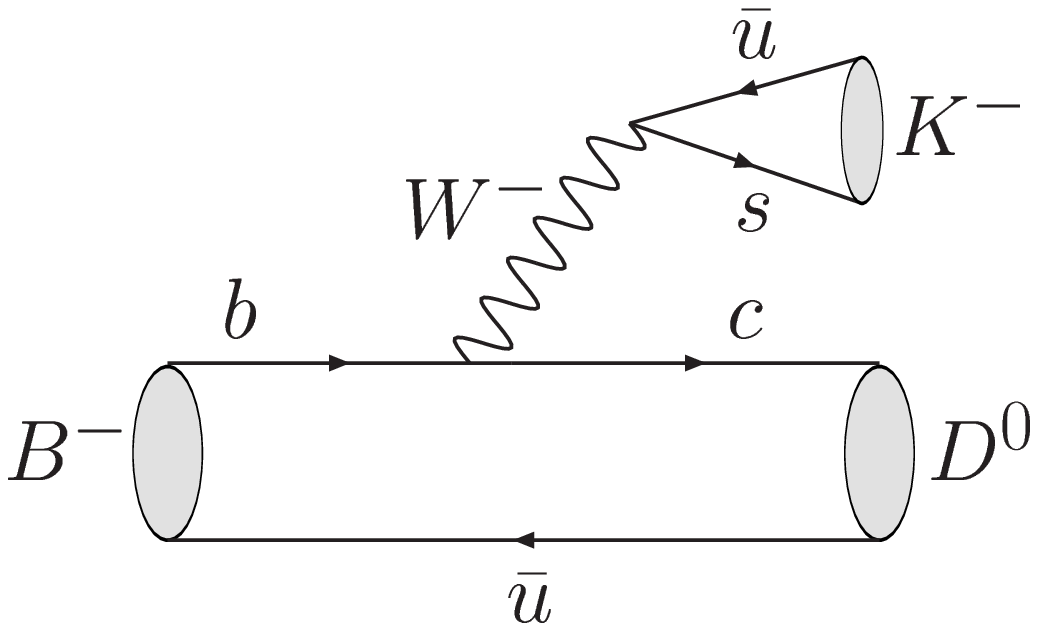}
\includegraphics[width=5cm]{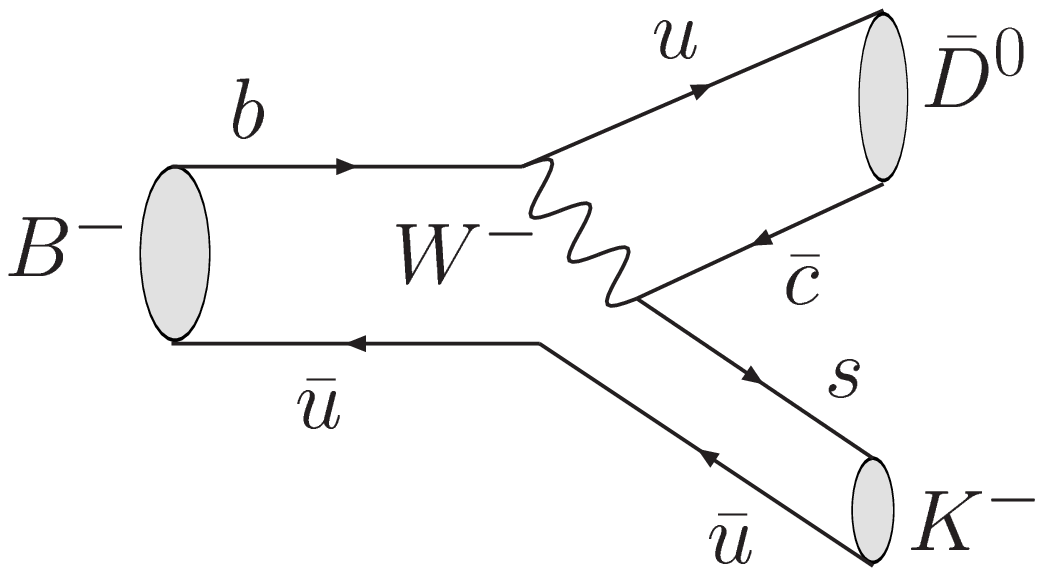}
\end{center}
\caption{Main Feynman diagrams contributing to the $B^-\to DK^-$ decay. The left decay proceeds through a $b\to c\bar u s$ transition, while the right diagram proceeds via a $b\to u\bar c s$ transition and is both color- and Cabibbo-suppressed.}
\label{fig:btodk_diagrams}
\end{figure}

In the following we will describe and discuss the main techniques to measure the angle $\gamma$ based on the measurement of $B\to DK$ decays and we will summarize the results.
 
\section{Measurement of $\gamma$ with flavor-tagged $B\to DK$ decays}
The most powerful method used so far exploits the interference between $b\to c\bar u s$ and $b\to u\bar c s$ amplitudes in $B\to DK$ decays, whose relative weak phase is $\gamma$. In charged $B$ decays the interference is between the $B^-\to D^0 K^-$ amplitude and the color- and Cabibbo-suppressed $B^-\to \bar D^0 K^-$ amplitude, when $D^0$ and $\bar D^0$ decay to a common final state. The leading interfering diagrams are shown in Fig.~\ref{fig:btodk_diagrams}. 

Let us introduce the decay amplitudes 
$A(B^-\rightarrow D^0 K^-)=A_c e^{i\delta_c}$, $A(B^-\rightarrow \bar{D}^0 K^-)=A_u e^{i(\delta_u-\gamma)}$, $A(D^0\rightarrow f)=A_fe^{i\delta_f}$ and $A(\bar D^0\rightarrow f)=A_{\bar{f}}e^{i\delta_{\bar{f}}}$, where $f$ is a generic final state of the $D$ meson. The parameters $\delta_c$, $\delta_u$, $\delta_f$ and $\delta_{\bar{f}}$ are strong phases, and $\gamma$ is the weak phase difference between $B^-\to D^0 K^-$ and $B^-\to\bar{D}^0 K^-$. $A_c$, $A_u$, $A_f$ and $A_{\bar{f}}$ are real and positive.
The decay amplitude of $B^-\to [f]_D K^-$ is
\begin{equation}\label{eq:amp2}
A(B^-\rightarrow [f]_DK^-) =A_c A_fe^{i(\delta_c+\delta_f)}+A_uA_{\bar{f}}e^{i(\delta_u+\delta_{\bar{f}}-\gamma)}\, ,
\end{equation}
and the rate can be written as

\begin{equation}
\Gamma(B^-\rightarrow [f]_DK^-)=A_c^2A_{\bar{f}}^2\left(A_f^2/A_{\bar{f}}^2 +r_B^2 +2r_BA_f/A_{\bar{f}}\re(e^{i(\delta_B+\delta_D-\gamma)})\right)\, ,\label{eq:rate2}
\end{equation}
where $r_B=A_u/A_c$, $\delta_B=\delta_u-\delta_c$ and
$\delta_D=\delta_{\bar{f}}-\delta_f$. The rate of the charge-conjugate mode
is obtained from Eq.~\ref{eq:rate2} by replacing $\gamma$ with $-\gamma$. 
Since with excellent approximation there are no other contributions besides the ones in Fig.~\ref{fig:btodk_diagrams}, $\gamma$ and the unknown hadronic parameters $r_B$ and $\delta_B$ can be constrained in a theoretically clean way by measuring the yields of $B^+$ and $B^-$, provided that the $D$ decays are chosen appropriately.
The same concept and equations apply to $B^\pm\to D^{*0}K^\pm$, $B^\pm \to D^0K^{*\pm}$ and to the flavor-tagged $B^0\to D^0 K^{*0}$ decays with small, though important, modifications\cite{bondar_gershon, btodkstar_gronau}\footnote{The decay $B\to DK^*$ can also be measured as part of the Dalitz plot analysis of $B\to D K\pi$. See~\cite{gershon} and references therein.}. The parameters $r_B$ and $\delta_B$ are not expected to have the same value in different $B\to D^{(*)0}K^{(*)}$ decays. The magnitude of $r_B$ is crucial because it measures the interference which allows the extraction of $\gamma$. For charged $B$ decays $r_B\approx c_F |V_{cs}V_{ub}^*/V_{us}V_{cb}^*|\sim 0.1-0.2$, where $c_F\sim 0.2-0.4$ is a color suppression factor. For neutral $B$ decays $r_B$ can be as large as 0.4, but the typical yields are significantly smaller making these decays less competitive at present.    

In the next three sections the main experimental methods explored so far are briefly discussed.

\subsection{The GLW method}\label{sec:glw}
In the method proposed by Gronau, London and Wyler (GLW)~\cite{glw} the $D$ meson is reconstructed to $CP$ eigenstate final states $f_{CP\pm}$, such as $K^+K^-$ ($CP=+1$) or $K^0_S\pi^0$ ($CP=-1$). Therefore $A_f/A_{\bar{f}}=1$, $\delta_D=0,\pi$ for $CP=1, -1$, and Eq.~\ref{eq:rate2} becomes
\begin{equation}\label{eq:glw1}
\Gamma(B^-\rightarrow [f_{CP\pm}]_DK^-)=A_c^2A_{f_{CP\pm}}^2(1+r_B^2\pm 2r_B\cos(\delta_B-\gamma))\, .
\end{equation}
Equation~\ref{eq:glw1} and its $CP$-conjugate are used to define four observables which depend on the physical parameters $\gamma$, $\delta_B$ and $r_B$:
\begin{eqnarray}
&&A_{CP\pm}=\frac{\Gamma(B^-\rightarrow D^0_{CP\pm}K^-)-\Gamma(B^+\rightarrow
    D^0_{CP\pm}K^+)}{\Gamma(B^-\rightarrow D^0_{CP\pm}K^-)+\Gamma(B^+\rightarrow
    D^0_{CP\pm}K^+)}=\frac{\pm2r_B\sin\delta_B\sin\gamma}{1+r_B^2\pm2 r_B\cos\delta_B\cos\gamma}\label{eq:glw2}\, ,\\
&&R_{CP\pm}=\frac{\Gamma(B^-\rightarrow D^0_{CP\pm}K^-)+\Gamma(B^+\rightarrow D^0_{CP\pm}K^+)}{2\Gamma(B^-\rightarrow D^0 K^-)}=1+r_B^2\pm2 r_B\cos\delta_B\cos\gamma\label{eq:glw3}\, ,
\end{eqnarray}
where $\Gamma(B^-\to D^0_{CP\pm}K^-)\equiv\Gamma(B^-\rightarrow [f_{CP\pm}]_DK^-)/BF(D^0\to f_{CP\pm})$ and $\Gamma(B^-\to D^0K^-)\equiv A_c^2\simeq\Gamma(B^-\to [K^-\pi^+]_DK^-)/ BF(D^0\to K^-\pi^+)$. $A_{CP\pm}$ and $R_{CP\pm}$ are bound by the relation $R_{CP+}A_{CP+}+R_{CP-}A_{CP-}=0$.

 Both BaBar and Belle have reconstructed $B^-\to D^0K^-$ and $B^-\to D^{*0}K^-$ decays with $D^{*0}\to D^0\pi^0$. BaBar has also selected the mode $D^{*0}\to D^0\gamma$ and the decay $B^-\to D^0 K^{*-}$ with $K^{*-}\to K_S^0\pi^-$. $D^0$ mesons have been reconstructed in $CP$-even ($K^+K^-$ and $\pi^+\pi^-$) and $CP$-odd ($K_S^0\pi^0$, $K_S^0\phi$ and $K_S^0\omega$) eigenstates. Recently also CDF has measured $B^-\to D^0K^-$ with $D^0\to K^+K^-$ and $\pi^+\pi^-$. The data samples used by BaBar and Belle consist of $382$ and $275$ million $B\bar B$ pairs, respectively, while CDF has used 1.0~fb$^{-1}$ of data. The results are summarized in Table~\ref{tab:glw}.

Due to the discrete 8-fold ambiguity in the extraction of $\gamma$ from $R_{CP\pm}$ and $A_{CP\pm}$, and the still large uncertainty on $r_B$, the GLW measurements have a poor constraining power on $\gamma$ when they are considered alone. However, they generally improve the knowledge of $r_B$, $\gamma$ and $\delta_B$ when combined with the results of the Dalitz method discussed in Sec.~\ref{sec:dalitz}. This aspect is further discussed in Sec.~\ref{sec:cart}.

\begin{table}[!h]
\begin{center}
\footnotesize
 \caption{
    \label{tab:glw}
 Summary of $R_{CP\pm}$ and $A_{CP\pm}$ measurements.}
\begin{tabular}{|c|c|c|c|c|c|}
\hline
 Mode & Experiment & $A_{CP+}$ & $A_{CP-}$ & $R_{CP+}$ & $R_{CP-}$ \\
\hline
$B\to D^0K$ & BaBar~\cite{glwbtodk_babar} & $0.27\pm 0.09\pm 0.04$ & $−0.09\pm 0.09\pm 0.02$ & $1.06\pm 0.10\pm 0.05$ & $1.03\pm 0.10\pm 0.05$\\
            & Belle~\cite{glwbtodk_belle} & $0.06\pm 0.14\pm 0.05$ & $−0.12\pm 0.14\pm 0.05$ & $1.13\pm 0.16\pm 0.08$ & $1.17\pm 0.14\pm 0.14$ \\
            & CDF~\cite{glwbtodk_cdf} & $0.39\pm 0.17\pm 0.04$ & --- & $1.30\pm 0.24\pm 0.12$  & --- \\
\hline
$B\to D^{*0}K$ & BaBar~\cite{glwbtodsk_babar} & $−0.11\pm 0.09\pm 0.01$ & $0.06\pm 0.10\pm 0.02$  & $1.31\pm 0.13\pm 0.03$ & $1.09\pm 0.12\pm 0.04$ \\
              & Belle~\cite{glwbtodk_belle} & $−0.20 \pm 0.22 \pm 0.04$ & $0.13 \pm 0.30 \pm 0.08$ & $1.41 \pm 0.25 \pm 0.06$ & $1.15 \pm 0.31 \pm 0.12$ \\
 \hline
$B\to D^0K^*$ & BaBar~\cite{glwbtodks_babar} & $0.09 \pm 0.13 \pm 0.06$ & $-0.23 \pm 0.21 \pm 0.07$ & $2.17 \pm 0.35 \pm 0.09$ & $1.03 \pm 0.27 \pm 0.13$ \\
\hline
  \end{tabular}
\end{center}
\end{table}

%
%
\subsection{The ADS method}
In the method proposed by Atwood, Dunietz and Soni (ADS)~\cite{ads} the $D$ meson is reconstructed in doubly Cabibbo-suppressed (DCS) states. We consider  $D^0\to K^+\pi^-$ as an example in the following discussion. The decay rate of the process $B^-\to [K^+\pi^-]_DK^-$ is the result of the interference between $B^-\to D^0K^-$ followed by the DCS $D^0\to K^+\pi^-$, and the suppressed $B^-\to \bar D^0 K^-$ followed by the Cabibbo-allowed $\bar D^0\to K^+\pi^-$.  From Eq.~\ref{eq:rate2} we find
\begin{equation}
\frac{\Gamma(B^\mp\rightarrow [K^\pm\pi^\mp]_DK^\mp)}{\Gamma(B^\mp\rightarrow [K^\mp\pi^\pm]_DK^\mp)}=r_B^2+r_D^2+2r_Br_D\cos(\delta_B+\delta_D\mp\gamma)\, ,\label{eq:ads_master}
\end{equation}
where both $r_D=A_f/A_{\bar{f}}=|A(D^0\rightarrow
K^+\pi^-)/A(D^0\rightarrow K^-\pi^+)|$ and the strong phase difference $\delta_D$ are measured independently~\cite{pdg2008,deltad_kpi}.
 Defining $R_{ADS}$ and $A_{ADS}$ as
\begin{eqnarray}
&&R_{ADS}=\frac{\Gamma(B^-\rightarrow [K^+\pi^-]_DK^-)+\Gamma(B^+\rightarrow [K^-\pi^+]_DK^+)}{\Gamma(B^-\rightarrow [K^-\pi^+]_D K^-)+\Gamma(B^+\rightarrow [K^+\pi^-]_D K^+)}\, ,\label{eq:ads1}\\
&&A_{ADS}=\frac{\Gamma(B^-\rightarrow [K^+\pi^-]_DK^-)-\Gamma(B^+\rightarrow [K^-\pi^+]_DK^+)}{\Gamma(B^-\rightarrow [K^+\pi^-]_D K^-)+\Gamma(B^+\rightarrow [K^-\pi^+]_D K^+)}\label{eq:ads2}
\end{eqnarray}
it follows
\begin{eqnarray}
&&R_{ADS}=r_B^2+r_D^2+2r_B\,r_D\cos\gamma\cos(\delta_B+\delta_D)\label{eq:rads}\, ,\\
&&A_{ADS}=2r_B\,r_D\sin\gamma\sin(\delta_B+\delta_D)/R_{ADS}\, .\label{eq:aads}
\end{eqnarray}
Since $r_D(K\pi)=(5.78\pm 0.08)\%$~\cite{hfag} and $r_B$ is expected to be around 10\%, the interference effect can be quite large.
Similar relations, with small modifications, are derived for $B^-\to D^{*0}K^-$ and $B^-\to D^0 K^{*-}$ decays (see~\cite{bondar_gershon} and refs. in Table~\ref{tab:ads}) and for multi-body $D^0$ final states~\cite{deltad_multibody,adsbtodk_kpipi0_babar}.
Both BaBar and Belle have reconstructed the decay $B^-\to D^0 K^-$ followed by $D^0\to K^+\pi^-$ on datasets of 467 and 657 million $B\bar B$ pairs, respectively. BaBar has also selected $B^-\to D^{*0} K^-$ with $D^{*0}\to D^0\pi^0$ and $D^{*0}\to D^0\gamma$ ($467\times 10^6\, B\bar B$), and $B^-\to D^0 K^{*-}$ with $K^{*-}\to K^0_S\pi^-$ ($379\times 10^6\, B\bar B$). On a dataset of 465~million $B\bar B$ pairs BaBar has performed the first measurement of flavor-tagged decays $B^0\to D^0 K^{*0}$ with $K^{*0}\to K^+\pi^-$, selected in the $D^0$ final states $K^+\pi^-$, $K^+\pi^-\pi^0$ and $K^+\pi^-\pi^-\pi^+$. The results are summarized in Table~\ref{tab:ads}. Due to the smallness of the involved branching fractions no evidence of signal has been observed so far and the null measurements have been used to set upper limits on $r_B$. The strongest hint of signal has been reported by BaBar with a statistical significance of $2.6\sigma$ for $B^\mp\to D^0[K^\pm\pi^\mp]K^\mp$. Even when a signal is observed the constraining power on $\gamma$ of the ADS method is weak when it is used alone, but it becomes
significant when the ADS information is combined with the other methods. This aspect is discussed in Sec.~\ref{sec:cart}.

\begin{table}[!h]
\begin{center}
\footnotesize
 \caption{
    \label{tab:ads}
 Summary of $R_{ADS}$ and $A_{ADS}$ measurements, and limits on $r_B$.}
\renewcommand{\arraystretch}{1.1}
\begin{tabular}{|l|c|c|c|c|}
\hline
 Mode & Experiment & $R_{ADS}[10^{-2}]$ & $A_{ADS}$ & $r_B$\\
\hline
$B^-\to D^0K^-$ & & & &\\
\hspace{0.3cm}$D^0\to K^+\pi^-$ & BaBar~\cite{adsbtodk_babar} & $1.36\pm 0.55\pm 0.27$ & $-0.70\pm 0.35^{+0.09}_{-0.14}$ & $ [0.09,0.193]$ @ 95\% CL \\
                  & Belle~\cite{adsbtodk_belle} & $0.8\pm 0.6 ^{+0.2}_{-0.3}$ & $-0.13^{+0.97}_{-0.88}\pm 0.26$  & $<0.19$ @ 90\% CL\\
\hspace{0.3cm}$D^0\to K^+\pi^-\pi^0$ &BaBar~\cite{adsbtodk_kpipi0_babar}& $1.2\pm 1.2\pm 0.9$ & --- & $<0.19$ @ 95\% CL\\
\hline
$B^-\to D^{*0}[D^0\pi^0]K^-$ & & & & \\
\hspace{0.3cm}$D^0\to K^+\pi^-$ & BaBar~\cite{adsbtodk_babar} & $1.76\pm 0.93\pm 0.42$ & $0.77\pm 0.35\pm 0.12$ & \\
$B^-\to D^{*0}[D^0\gamma]K^-$ & & & &$ [0.007,0.176]$ @ 95\% CL\\
\hspace{0.3cm}$D^0\to K^+\pi^-$ & BaBar~\cite{adsbtodk_babar} & $1.3\pm 1.4\pm 0.5$ & $0.36\pm 0.94^{+0.25}_{-0.41}$ &\\
\hline
$B^-\to D^0K^{*-}$ & & & &\\
\hspace{0.3cm}$D^0\to K^+\pi^-$ & BaBar~\cite{glwbtodks_babar} & $6.6\pm 3.1\pm 1.0$ & $-0.34 \pm 0.43 \pm 0.16$ & $ [0.17,0.43]$ @ 95\% CL\\
 & & & & \scriptsize {(combined with GLW)}\\
\hline
$B^0\to D^0K^{*0}$ & & & & \\
\hspace{0.3cm}$D^0\to K^+\pi^-$ & BaBar~\cite{adsb0todks_babar} & $6.7^{+7.0}_{-5.4}\pm 1.8$ & --- &  \\
\hspace{0.3cm}$D^0\to K^+\pi^-\pi^0$ & BaBar~\cite{adsb0todks_babar} & $6.0^{+5.5}_{-3.7}\pm 0.9$ & --- & $[0.07,0.41]$ @ 95\% CL\\
\hspace{0.3cm}$D^0\to K^+\pi^-\pi^-\pi^+$ & BaBar~\cite{adsb0todks_babar} & $13.7^{+11.3}_{-9.5}\pm 2.2$ & --- & \\
\hline
  \end{tabular}
\end{center}
\end{table}

%
%
\subsection{The Dalitz or GGSZ method}\label{sec:3body}\label{sec:dalitz}
If $D^0$ decays to a 3-body final state such as $D\to K^0_S\pi^+\pi^-$, the decay amplitudes of $D^0$ and $\bar D^0$ can be written as $A_fe^{i\delta_f}=f(m_-^2,m_+^2)$ and $A_{\bar{f}}e^{i\delta_{\bar{f}}}=f(m_+^2,m_-^2)$, where $m_-^2$ and $m_+^2$ are the squared masses of the $K^0_S\pi^-$ and $K^0_S\pi^+$ combinations. 
The rate in Eq.~\ref{eq:rate2} becomes
\begin{eqnarray}\label{eq:dalitzrate1}
&\Gamma(B^\mp\to [K^0_S\pi^-\pi^+]_DK^\mp)\propto& |f(m_\mp^2,m_\pm^2)|^2+r_B^2|f(m_\pm^2,m_\mp^2)|^2+ \\
&&2r_B|f(m_\mp^2,m_\pm^2)||f(m_\pm^2,m_\mp^2)|\cos(\delta_B+\delta_D(m_\mp^2,m_\pm^2)\mp\gamma)\nonumber
  \, ,
\end{eqnarray}
where $\delta_D(m_\mp^2,m_\pm^2)$ is the strong phase difference
between $f(m_\pm^2,m_\mp^2)$ and $f(m_\mp^2,m_\pm^2)$. The amplitudes $f(m_\pm^2,m_\mp^2)$ are measured through a Dalitz plot analysis on a large sample of flavor-tagged $D^0$ decays. Therefore, the $B^\pm$ yields in Eq.~\ref{eq:dalitzrate1} only depend on the unknowns $\gamma$, $\delta_B$ and $r_B$. The great advantage of this method~\cite{dalitz_method} is that the relation between the signal yields and the physics parameters varies over the Dalitz plot, making possible the extraction of $\gamma$ with only a 2-fold ambiguity ($\gamma\to\gamma+180^\circ$). Furthermore, the $D^0\to K^0_S\pi^+\pi^-$ branching fractions is relatively large ($\sim 3\%$).
Since the direct extraction of $r_B$, $\delta_B$ and $\gamma$ through a maximum likelihood fit (MLF) using Eq.~\ref{eq:dalitzrate1} overestimates $r_B$ and underestimates the statistical error of $\gamma$ and $\delta_B$, it is convenient to express Eq.~\ref{eq:dalitzrate1} in terms of the cartesian coordinates $x_\pm=\re[r_B e^{i(\delta_B\pm\gamma)}]$, $y_\pm=\im[r_B e^{i(\delta_B\pm\gamma)}]$,
\begin{equation}\label{eq:dalitzrate2}
\Gamma(B^\mp\to D^0[\to K^0_S\pi^+\pi^-]K^\mp)\propto
|f_{\mp}|^2+(x^2_\mp+y^2_\mp)|f_{\pm}|^2+2\left[x_\mp \re[f_{\mp}f_{\pm}^*]+y_\mp \im[f_{\mp}f_{\pm}^*]\right]
\end{equation}
where the notation was simplified using $f_{\pm}=f(m_\pm^2,m_\mp^2)$. The extraction of $x_\pm$ and $y_\pm$ with a MLF using Eq.~\ref{eq:dalitzrate2} is unbiased. The physics parameters $\gamma$, $r_B$ and $\delta_B$ are extracted from $x_\pm$ and $y_\pm$ with a frequentist statistical procedure.

BaBar has used this approach to measure the angle $\gamma$ with $B^-\rightarrow D^{0} K^-$, $D^{*0}K^-$ ($D^{*0}\to D^0\pi^0$ and $D^0\gamma$) and $D^0K^{*-}$ ($K^{*-}\to K^0_S\pi^-$), with $D^0\rightarrow K^0_s\pi^+\pi^-$ and $K^0_SK^+K^-$ using a sample of 386 million $B\bar B$ pairs\cite{dalitz_babar}. Belle has selected $B^-\to D^{(*)0}K^-$ decays ($D^{*0}\to D^0\pi^0$ and $D^0\gamma$) with $D^0\to K^0_S\pi^+\pi^-$ on a sample of 657 million $B\bar B$ pairs~\cite{dalitzbtodk_belle}. The $D^0\to K^0_S\pi^+\pi^-$ and $D^0\to K^0_SK^+K^-$ decay amplitudes are determined with Dalitz plot analyses of large and very pure samples of $D^0$ mesons from $D^{*+}\to D^0\pi^+$ decays produced in $e^+e^-\to c\bar c$ events. The amplitudes are described using an isobar model, consisting of a coherent sum of two-body amplitudes (parameterized using relativistic Breit-Wigner) plus a ``nonresonant'' term. BaBar has described the $\pi\pi$ and $K\pi$ S-wave amplitudes in $D^0\to K^0_S\pi^+\pi^-$ using a $K$-matrix formalism (see~\cite{dalitz_babar} for a detailed discussion).

\begin{figure}[!h]
\begin{center}
\includegraphics[width=4.8cm]{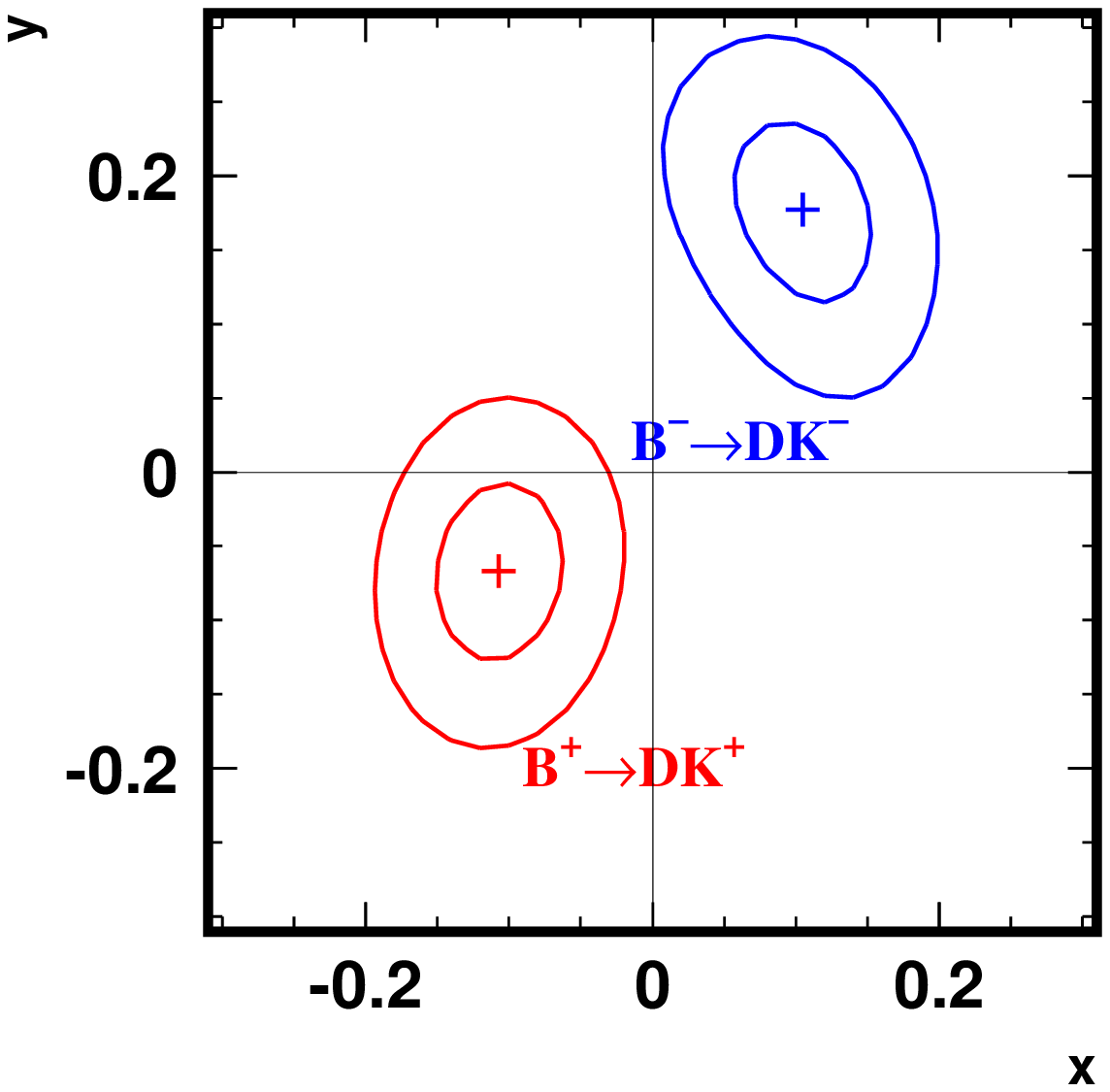}
\includegraphics[width=4.8cm]{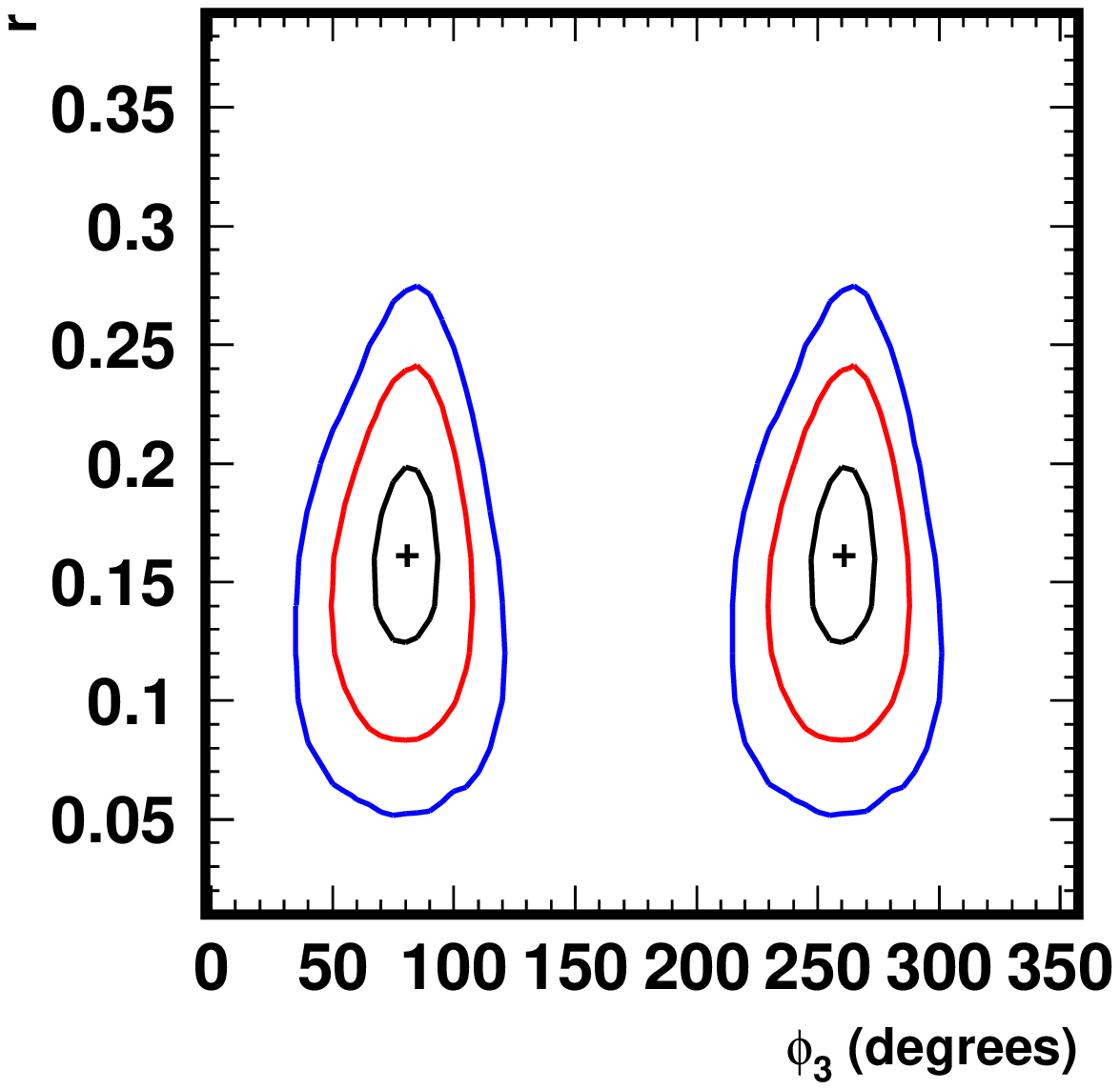}
\includegraphics[width=4.8cm]{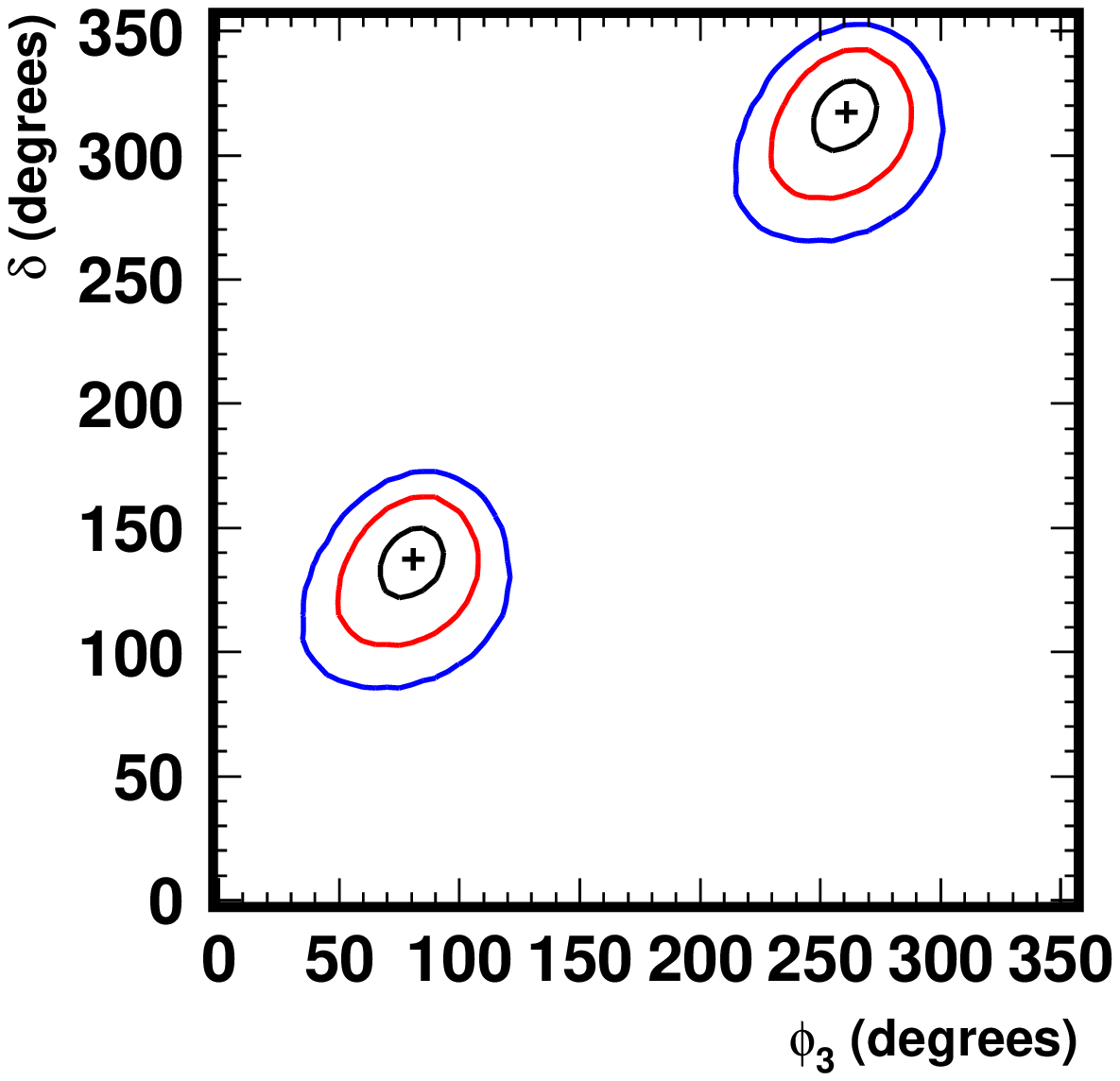}
\includegraphics[width=4.4cm,height=4.6cm]{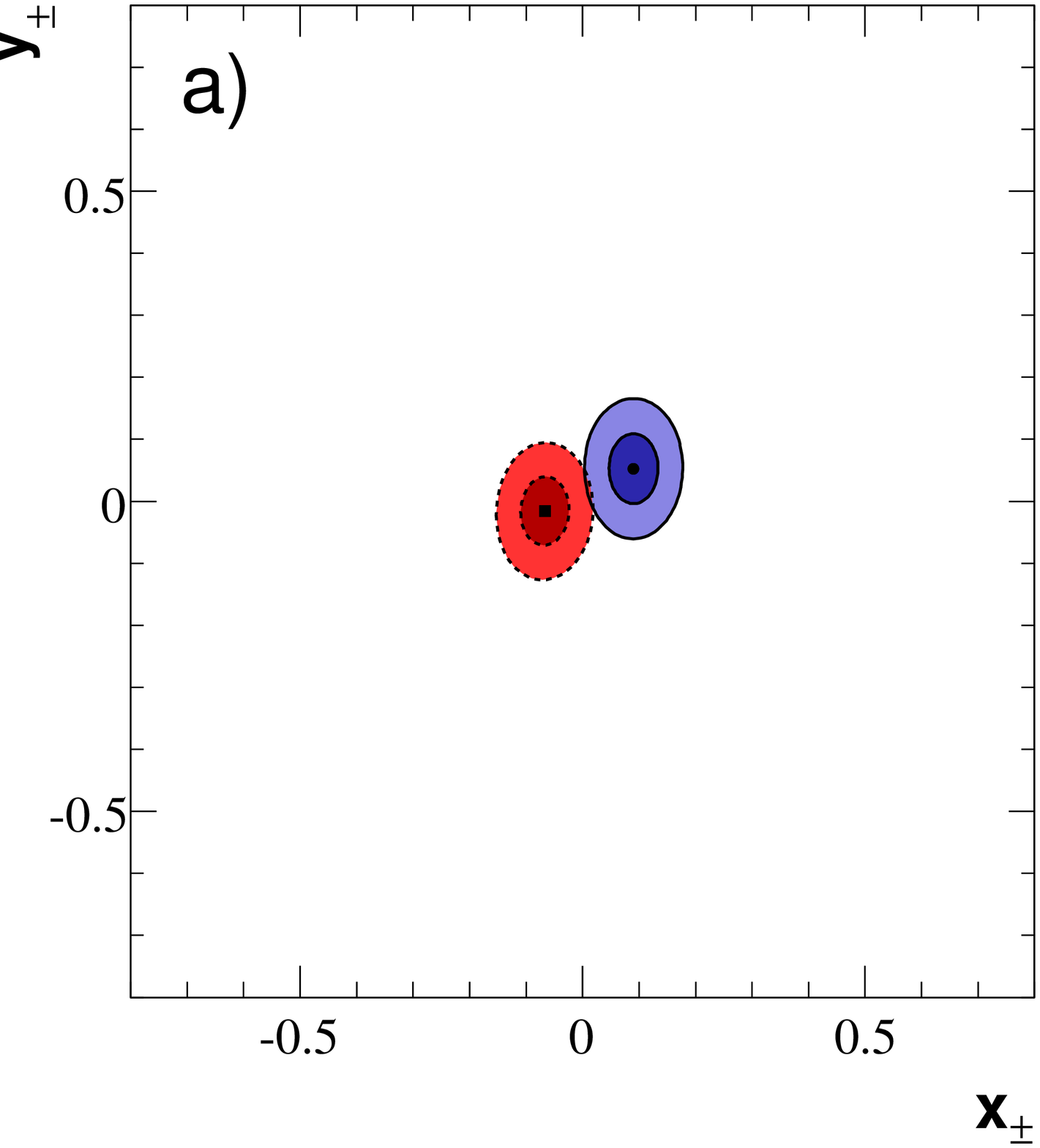}
\includegraphics[width=4.8cm]{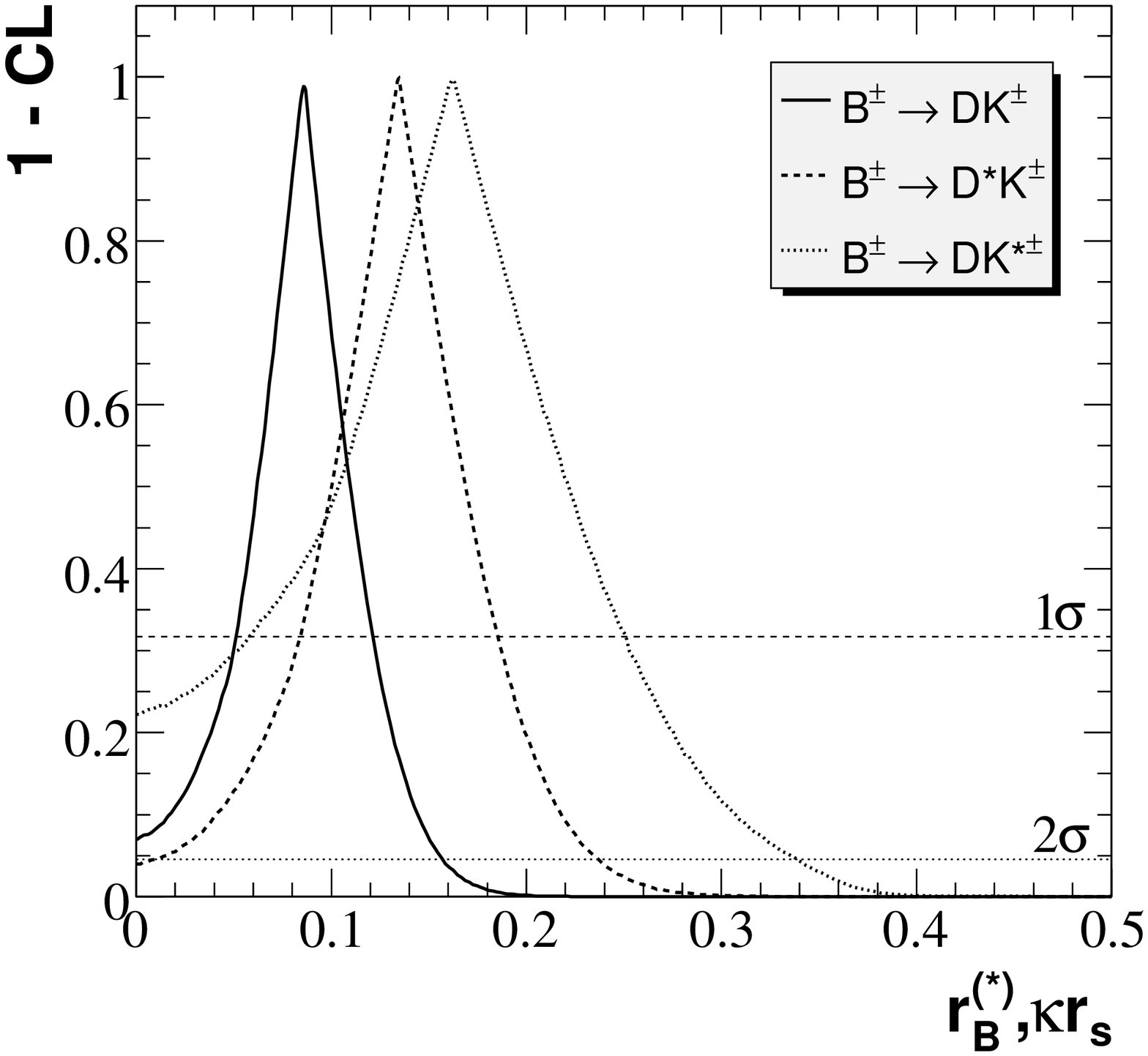}
\includegraphics[width=4.8cm]{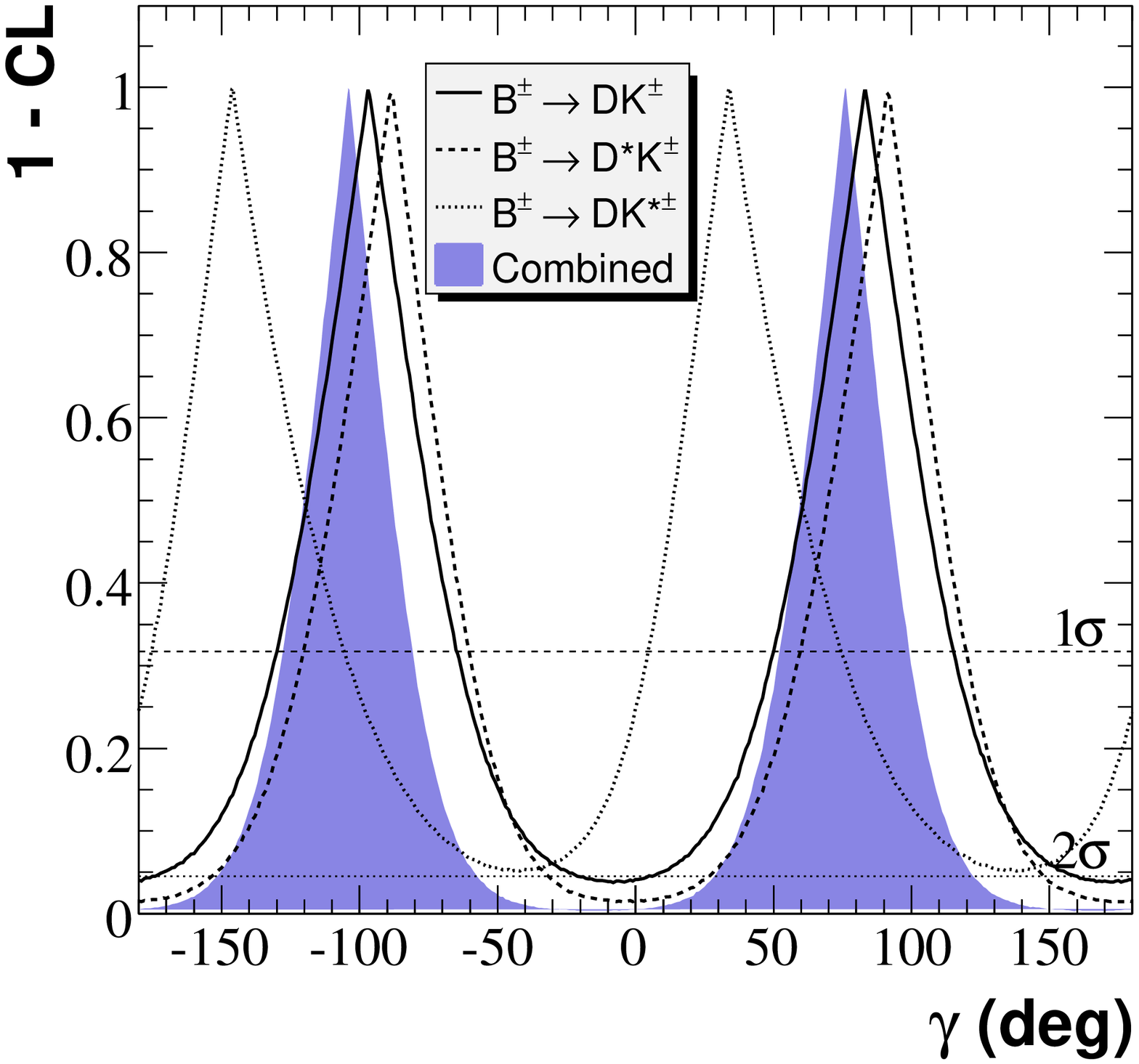}
\end{center}
\caption{Top row: Belle; bottom row: BaBar. 1$\sigma$ and 2$\sigma$ contours in the $(x_\pm,y_\pm)$ plane of $B^\pm\to D^0K^\pm$ (top-left); projections of confidence regions for the $B\to DK$ mode onto the $(r_B,\gamma)$ plane (top-center) and onto the $(\delta_B,\gamma)$ plane (top-right); 1- and 2-standard-deviation regions in the $(x_\pm,y_\pm)$ plane of $B^\pm\to D^0K^\pm$ (bottom-left); 1-CL as a function of $r_B$ (bottom-center) and $\gamma$ (bottom-right) for $B\to D^0K$, $D^{*0}K$ and $D^0K^*$.}
\label{fig:dalitz}
\end{figure}

The results for $x_\pm$ and $y_\pm$ are reported in Table~\ref{tab:dalitz}. From the $(x_\pm,y_\pm)$ confidence regions both BaBar and Belle determine the 1$\sigma$ confidence intervals of $\gamma$, $\delta_B$ and $r_B$ using a frequentist procedure. BaBar finds $\gamma=(76\pm 22\pm 5\pm 5)^\circ$ and Belle $\gamma=(76^{+12}_{-13}\pm 4\pm 9)^\circ$ ($B^\pm\to D^{(*)}K^\pm$), where the solution closest to the SM average has been quoted. All results are reported in Table~\ref{tab:dalitz_gamma}. Figure~\ref{fig:dalitz} shows the $(x_\pm,y_\pm)$ contours for $B\to D^0K$ as measured by Belle (top-left) and BaBar (bottom-left), the projections of confidence regions onto the $(\gamma,r_B)$ and $(\gamma,\delta_B)$ plane obtained by Belle, and the confidence-level as a function of $r_B$ and $\gamma$ found by BaBar. The combined $B\to D^{(*)0}K^{(*)}$ measurements of BaBar and Belle correspond to 3.0$\sigma$ and 3.5$\sigma$ evidence of $CP$ violation, respectively.

\begin{table}[!t]
\scriptsize
 \caption{
    \label{tab:dalitz}
 Summary of $x_\pm$ and $y_\pm$ measurements. The third error is the systematic uncertainty associated to the Dalitz model of the $D$ final state.}
\renewcommand{\tabcolsep}{0.1cm}
\begin{tabular}{|c|c|c|c|c|c|}
\hline
{\fns Mode} & {\fns Experiment} & {\fns$x_+$} $[10^{-2}]$& {\fns$y_+$} $[10^{-2}]$ & {\fns $x_-$} $[10^{-2}]$& {\fns $y_-$} $[10^{-2}]$\\
\hline
$B\to D^0K$ & & & & &\\
\hspace{0.3cm}Dalitz& BaBar~\cite{dalitz_babar} & $-6.7 \pm 4.3 \pm 1.4 \pm 1.1$ & $-1.5 \pm 5.5 \pm 0.6 \pm 0.8$ & $9.0 \pm 4.3 \pm 1.5 \pm 1.1$  & $5.3 \pm 5.6 \pm 0.7 \pm 1.5$\\
                    & Belle~\cite{dalitzbtodk_belle,dalitzsys_belle_notes} & $-10.7 \pm 4.3 \pm 1.1 \pm 5.5$ & $-6.7 \pm 5.9 \pm 1.8 \pm 6.3$ & $10.5 \pm 4.7 \pm 1.1 \pm 6.4$ & $17.7 \pm 6.0 \pm 1.8 \pm 5.4$\\
\hspace{0.3cm}GLW& BaBar~\cite{glwbtodk_babar} & $-9\pm 5\pm 2$ & --- & $10\pm 5\pm 3$ & --- \\
\hline
$B\to D^{*0}K$ & & & & &\\
\hspace{0.3cm}Dalitz& BaBar~\cite{dalitz_babar} & $13.7 \pm 6.8 \pm 1.4 \pm 0.5$ & $8.0 \pm 10.2 \pm 1.0 \pm 1.2$ & $-11.1 \pm 6.9 \pm 1.4 \pm 0.4$ &$-5.1 \pm 8.0 \pm 0.9 \pm 1.0$ \\
                    & Belle~\cite{dalitzbtodk_belle,dalitzsys_belle_notes,dalitzbtodsk_belle_notes} & $13.3 \pm 8.3 \pm 1.8 \pm 8.1$ &$13.0 \pm 12.0 \pm 2.2 \pm 6.3$  & $2.4 \pm 14.0 \pm 1.8 \pm 9.0$ & $-24.3 \pm 13.7 \pm 2.2 \pm 4.9$\\
\hspace{0.3cm}GLW& BaBar~\cite{glwbtodsk_babar} & $11\pm 6\pm 2$ & --- & $0\pm 6\pm 2$ & --- \\
\hline
$B\to D^0K^*$ & & & & &\\
\hspace{0.3cm}Dalitz& BaBar~\cite{dalitz_babar} &$−11.3 \pm 10.7 \pm 2.8 \pm 1.8$  & $12.5 \pm 13.9 \pm 5.1 \pm 1.0$ & $11.5 \pm 13.8 \pm 3.9 \pm 1.4$ &$22.6 \pm 14.2 \pm 5.8 \pm 1.1$ \\
                    & Belle~\cite{dalitzbtodks_belle} & $-10.5^{+17.7}_{-16.7} \pm 0.6 \pm 8.8$ &$-0.4^{+16.4}_{-15.6} \pm 1.3 \pm 9.5$  & $-78.4^{+24.9}_{-29.5} \pm 2.9 \pm 9.7$ & $-28.1^{+44.0}_{-33.5} \pm 4.6 \pm 8.6$\\
\hspace{0.3cm}GLW& BaBar~\cite{glwbtodks_babar} & $18\pm 14\pm 5$  & ---  & $38\pm 14\pm 5$ & --- \\
\hline
\end{tabular}
\end{table}

\begin{table}[b]
\begin{center}
 \caption{
    \label{tab:dalitz_gamma}
Measurement of $\gamma$, $\delta_B$ and $r_B$ in $B^-\to D^{(*)0}K^{(*)-}$ decays reconstructed by BaBar and Belle.}
\renewcommand{\arraystretch}{1.2}
\begin{tabular}{|c|c|c|}
\hline
Parameter & BaBar & Belle\\
\hline
$\gamma$ & $76\pm 22\pm 5\pm 5$& $76^{+12}_{-13}\pm 4\pm 9$\\
$r_B(D^0K)$ & $0.086\pm 0.035\pm 0.010\pm 0.011$ & $0.161^{+0.040}_{-0.038}\pm 0.011\pm 0.049$\\
$\delta_B(D^0K)$ & $(109^{+28}_{-31}\pm 4\pm 7)^\circ$ & $(137.4^{+13.0}_{-15.7}\pm 4.0\pm 22.9)^\circ$ \\
$r_B(D^{*0}K)$ & $0.135\pm 0.051\pm 0.011\pm 0.005$ & $0.196^{+0.072}_{-0.069}\pm 0.012 ^{+0.062}_{-0.012}$\\
$\delta_B(D^{*0}K)$ & $(297^{+28}_{-30}\pm 5\pm 4)^\circ$&$(341.9^{+18.0}_{-19.6}\pm 3.0\pm 22.9)^\circ$ \\
$r_B(D^0K^*)$ & $0.163^{+0.088}_{-0.105}\pm 0.037\pm 0.021$& $0.564^{+0.216}_{-0.155}\pm 0.041\pm 0.084$ \\
$\delta_B(D^0K^*)$ &$(104^{+43}_{-41}\pm 17\pm 5)^\circ$ & $(242.6^{+20.2}_{-23.2}\pm 2.5\pm 49.3)^\circ$\\
\hline
\end{tabular}
\end{center}
\end{table}

\begin{figure}[!h]
\begin{center}
\includegraphics[width=7.2cm,height=6.6cm]{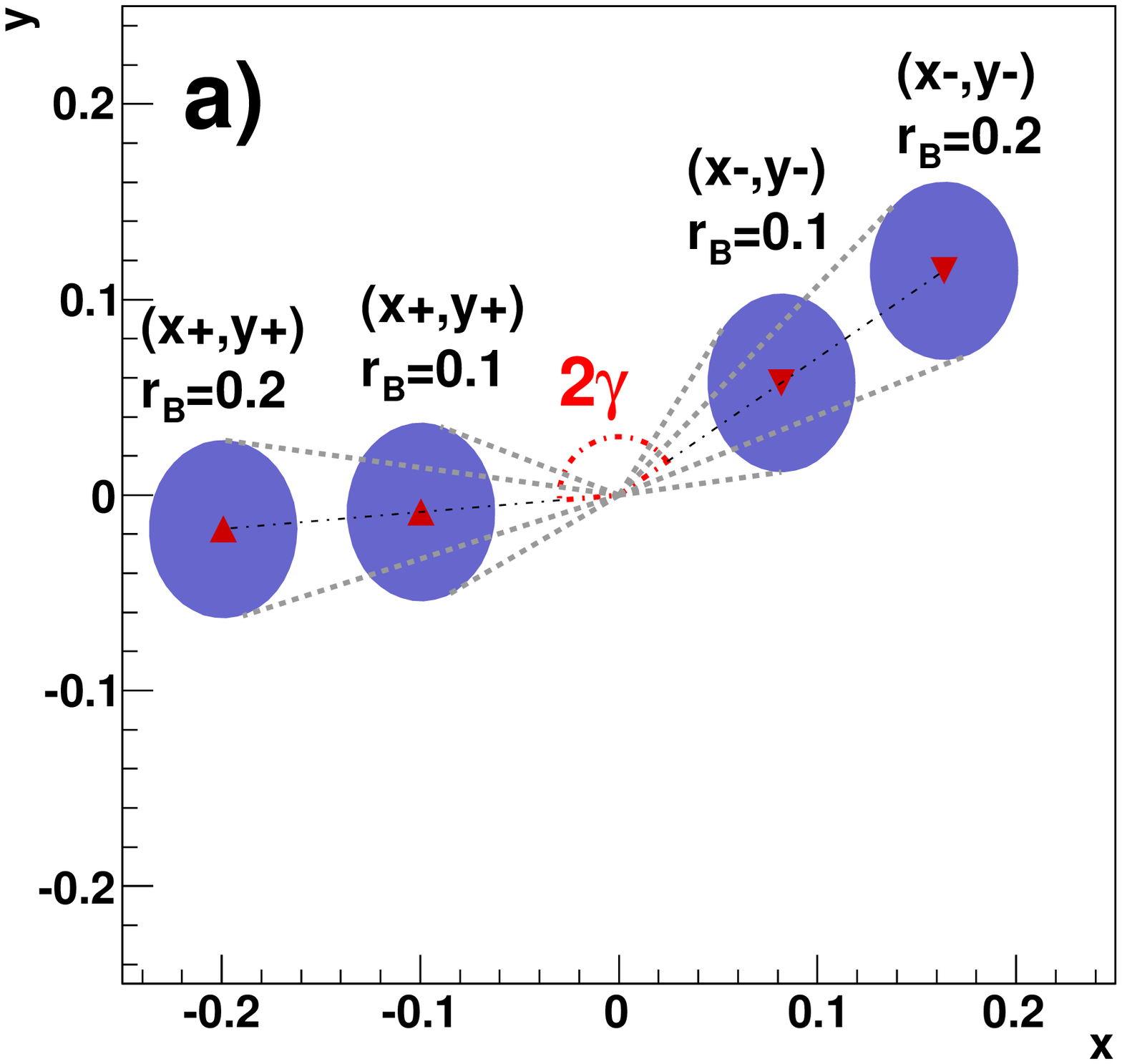}
\includegraphics[width=7.2cm]{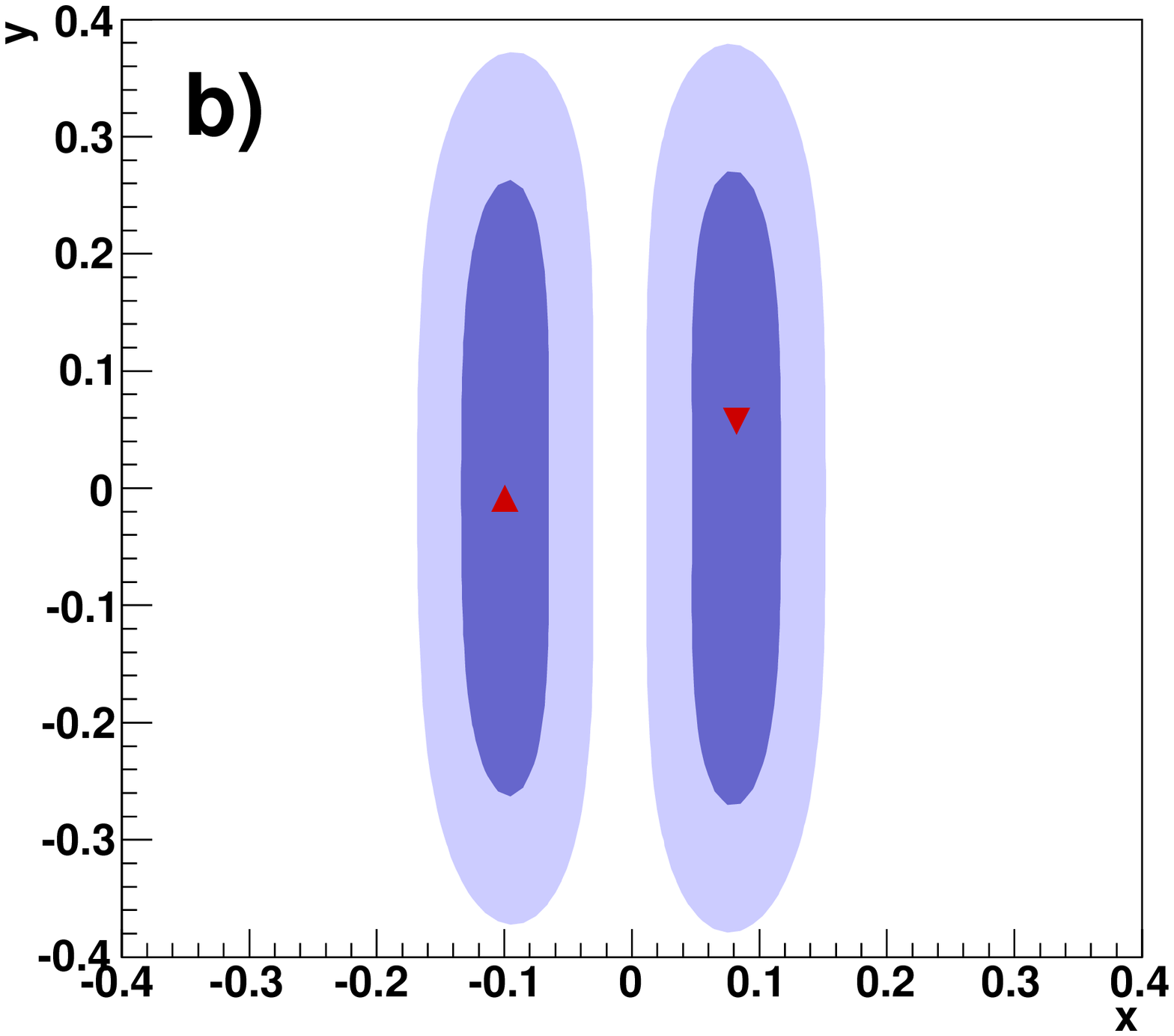}
\includegraphics[width=7.2cm]{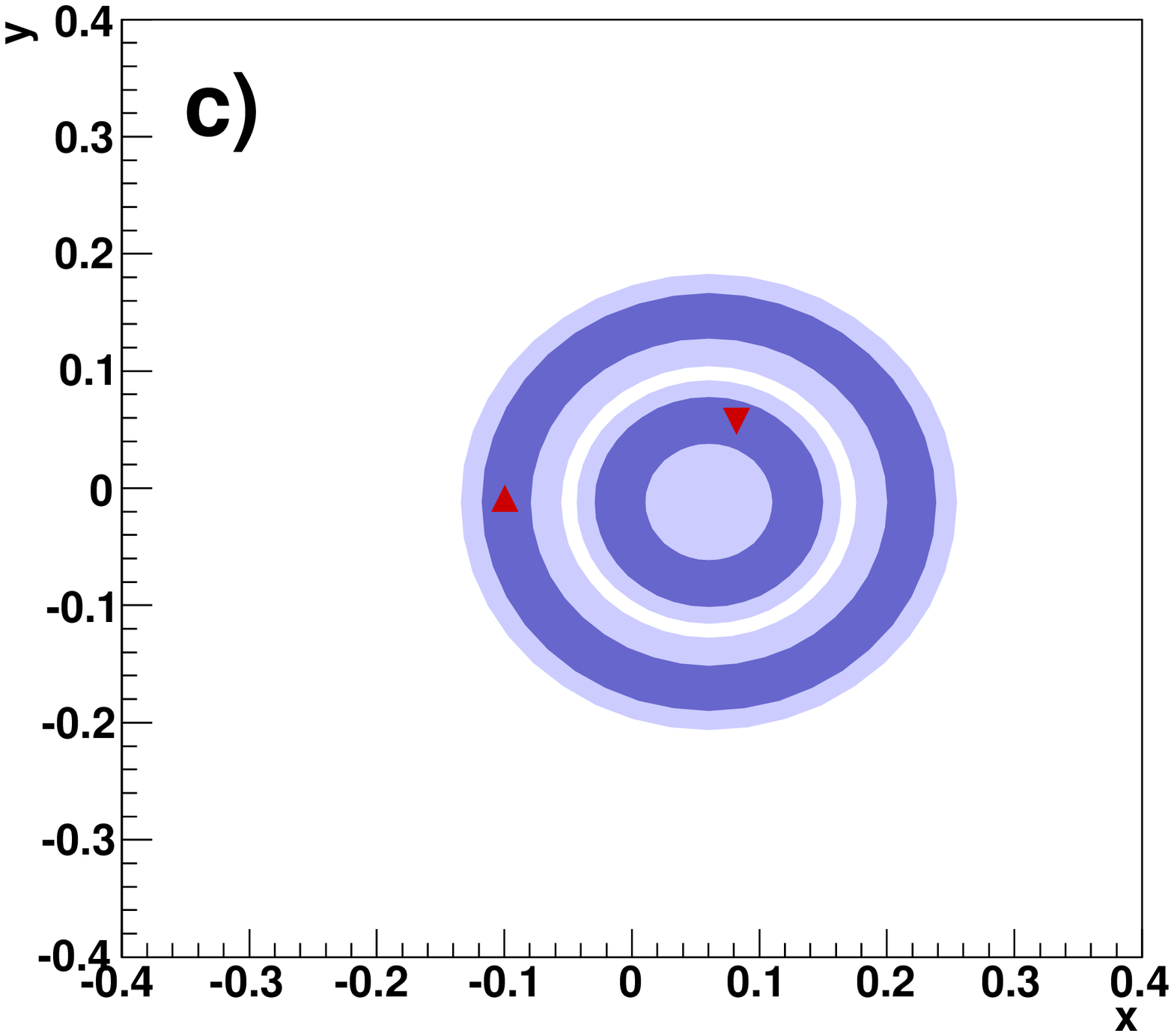}
\includegraphics[width=7.2cm]{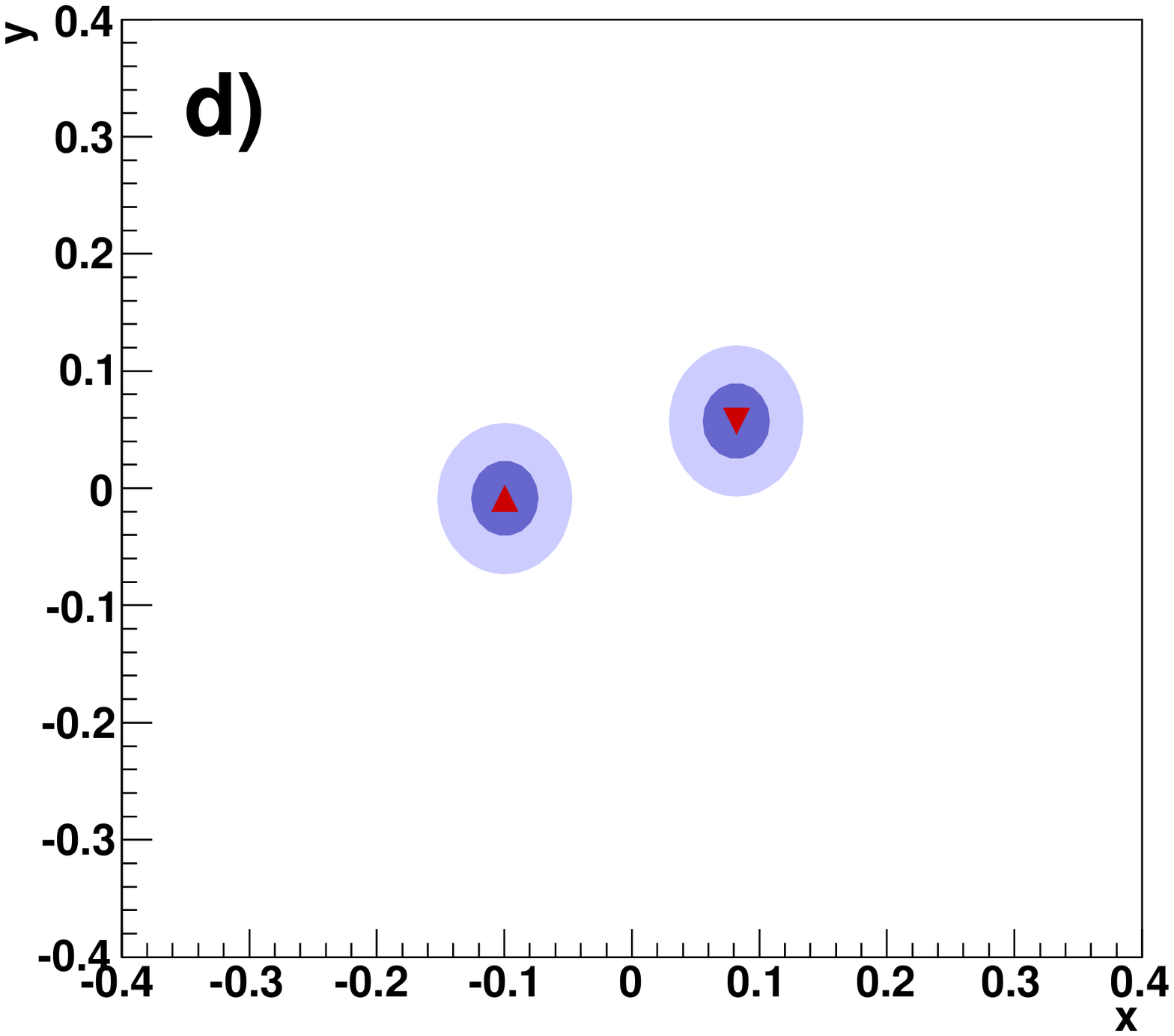}
\end{center}
\caption{a): Geometric definition of $\gamma$ in the $(x_\pm,y_\pm)$ plane and pictorial description of how its uncertainty changes as a function of $r_B$. Circles represent $x_\pm,y_\pm$ error regions for $B^+$ ($\blacktriangle$) and $B^-$ ($\blacktriangledown$) and the dashed lines delimit the range of variation of $2\gamma$. b,c,d): 1- and 2-standard deviation regions in the $(x_\pm,y_\pm)$ plane using the GLW (b), ADS (c) and Dalitz (d) methods, on a dataset of 1ab$^{-1}$. The assumptions used are discussed in the text.}\label{fig:xy_cons}
\end{figure}

It is interesting to ask why the statistical error measured by Belle is about twice smaller than what BaBar finds, despite the fact that the experimental observables $x_\pm,y_\pm$ have similar uncertainties. The error on $\gamma$ scales roughly as $1/r_B$. Since Belle measurements have central values of $r_B$ between 1.5 and 3.5 times larger than BaBar values, though consistent within the errors, the resulting $\gamma$ uncertainty of Belle is smaller. The concept is illustrated in Fig.~\ref{fig:xy_cons}a. Fluctuactions of the $\gamma$ error due to the ``$1/r_B$'' effect are expected to decrease as the relative uncertainty on $r_B$ will become smaller.

At present the Dalitz method has the best sensitivity to $\gamma$. However, the uncertainty associated to the Dalitz model of the $D$ final state is already the dominant contribution to the systematic error and it may be difficult to reduce it greatly in the future. 
To bypass this limit a model-independent analysis is required~\cite{dalitz_method}. At the price of a small loss of statistical power~\cite{bondar_poluetkov}, the method is free of model-dependent assumptions on the $D$ decay and therefore it is a promising approach to follow at LHCb and at the next generation $B$-factories. It requires the use of entangled $\Psi(3770)\to D\bar D$ decays at tau-charm factories such as CLEO-c, BES-III or next generation Super Flavor factories~\cite{cdr_superb}. It has been recently shown that using 818~pb$^{-1}$ of CLEO-c data an error on $\gamma$ of $\sim 2^\circ$ associated to the knowledge of the relative strong phase difference of $D^0\to K^0_S\pi^+\pi^-$ and $\bar D^0\to K^0_S\pi^+\pi^-$ can be obtained~\cite{cleoc}.
 
BaBar has applied the Dalitz method also to flavor-tagged $B^0\to D^0K^{*0}$ decays, with $K^{*0}\to K^+\pi^-$ and $D^0\to K^0_S\pi^+\pi^-$. On a dataset of 371 million $B\bar B$ pairs $39\pm 9$ signal candidates have been selected. Using the $D^0$ Dalitz plot model obtained in the charged $B$ analysis and imposing an external measurement of $r_B$, a loose constraint on $\gamma$ was set~\cite{babar_B0todks}. Self tagging $B^0\to D^0 K^{*0}$ decays and time-dependent measurements of $B^0\to D K^0_S$ decays are expected to be powerful tools to measure $\gamma$ at LHCb and at the next generation $B$-factories, where their production will be abundant and the large interference ($r_B\sim 0.4$) can be fully exploited.

\section{Note on the cartesian coordinates}\label{sec:cart}
The output of the GLW and ADS methods can be expressed in terms of the same cartesian coordinates $x_\pm=r_B\cos(\delta_B\pm\gamma)$ and $y_\pm=\sin(\delta_B\pm\gamma)$ measured with the Dalitz method. This alternative way to quote the results can be useful when different methods are compared or combined. In this section we will derive the alternative parameterization and will discuss some of the advantages with respect to the classic observables.

From the definitions in Sec.~\ref{sec:glw} we can write $\Gamma(B^-\to D^0_{CP\pm}K^-)/\Gamma(B^-\to D^0K^-)=1+x_-^2+y_-^2\pm 2x_-$, from which it follows
\begin{eqnarray}
&&\frac{1}{4}\left(\frac{\Gamma(B^-\to D^0_{CP+}K^-)}{\Gamma(B^-\to D^0K^-)}-\frac{\Gamma(B^-\to D^0_{CP-}K^-)}{\Gamma(B^-\to D^0K^-)}\right)=x_-\label{eq:glw_x1}\, ,\\
&&\frac{1}{2}\left(\frac{\Gamma(B^-\to D^0_{CP+}K^-)}{\Gamma(B^-\to D^0K^-)}+\frac{\Gamma(B^-\to D^0_{CP-}K^-)}{\Gamma(B^-\to D^0K^-)}\right)-1=x_-^2+y_-^2\label{eq:glw_x2}\, .
\end{eqnarray}

The same relations hold for $x_+$ and $x_+^2+y_+^2$ after replacing $B^-$ with $B^+$. Equation~\ref{eq:glw_x1} has been used by BaBar to measure $x_\pm$ with the GLW analysis\footnote{The relation was expressed in terms of $R_{CP\pm}$ and $A_{CP\pm}$.}, whose results are reported in Table~\ref{tab:dalitz}. It is interesting that the measurements of $x_\pm$ performed with the GLW and Dalitz methods have about the same uncertainty on datasets with similar size. The constraint given by Eq.~\ref{eq:glw_x2} is much looser because of the quadratic dependence on $x,y$ and the fact that $r_B\ll 1$. Therefore, the GLW method measures $x_\pm$ quite precisely but not $y_\pm$, and this is the reason why it can hardly constrain $\gamma$ when it is considered alone.  
When combined with the Dalitz method, however, the overall error of $\gamma$ can improve significantly.

Proceeding in a similar way, Eq.~\ref{eq:ads_master} of the ADS method can be written as
\begin{equation}
\frac{\Gamma(B^\mp\rightarrow [K^\pm\pi^\mp]_DK^\mp)}{\Gamma(B^\mp\rightarrow [K^\mp\pi^\pm]_DK^\mp)}=\left(x_\mp+r_D\cos\delta_D\right)^2+\left(y_\mp-r_D\sin\delta_D\right)^2  \, ,\label{eq:xy_ads}
\end{equation}
that represents two circles in the $(x_\pm,y_\pm)$ plane
centered at $(-r_D\cos\delta_D,\ r_D\sin\delta_D)$ and with radii $R_\mp=\sqrt{\frac{\Gamma(B^\mp\rightarrow [K^\pm\pi^\mp]_DK^\mp)}{\Gamma(B^\mp\rightarrow [K^\mp\pi^\pm]_DK^\mp)}}$. It is not possible to determine $\gamma$ with only the ADS analysis of $B\to D^0K$ because the true $x_\pm,y_\pm$ points are delocalized over two circles\footnote{In the case of $B\to D^{*0}K$ the circles associated to $D^0\pi^0$ and $D^0\gamma$ are centered at opposite points $(\pm r_D\cos\delta_D,\pm r_D\sin\delta_D)$ because of the effect pointed out in~\cite{bondar_gershon}; therefore the resulting constraint in the $(x_\pm,y_\pm)$ plane is the intersection of two circles and in principle it is possible to determine $\gamma$ up to a discrete ambiguity.}. However, the measurement can be combined with the Dalitz and GLW analyses to reduce the overall error of $x_\pm,y_\pm$, and therefore the uncertainty of $\gamma$. Figure~\ref{fig:xy_cons} shows the constraints provided by the GLW, ADS and Dalitz measurements with $B\to D^0 K$ decays on a dataset of 1ab$^{-1}$, assuming $\{\gamma,r_B,\delta_B,r_D,\delta_D\}=\{75^\circ,0.1,110^\circ,0.06,191^\circ\}$ in a scenario where the measured observables are centered to their true value. The uncertainty on the values of $r_D$ and $\delta_D$ was neglected when drawing the ADS constraint.

\section{Conclusions}
Despite the fact that BaBar and Belle have collected almost 1.5~ab$^{-1}$ of data, a precise measurement of the CKM phase $\gamma$ is not yet available. This is not surprising if one thinks that the typical branching fractions of the involved decays are of the order of $10^{-6}$ or smaller and the interference term is found to be around 10\% in the main decay modes. In fact, the sensitivity reached by the $B$-factories is much better than it was initially foreseen.
The decays $B^\pm\to D^{(*)0} K^{(*)\pm}$ are currently the most sensitive tool.
The method where the $D$ meson decays to $K^0_Sh^+h^-$ $(h=\pi,K)$ has the smallest uncertainty, ranging between $15^\circ$ and $24^\circ$ including systematics. The dominant source of systematic uncertainty currently comes from the Dalitz model of the $D$ final state. 

\begin{figure}[]
\begin{center}
\includegraphics[width=6cm,height=6cm]{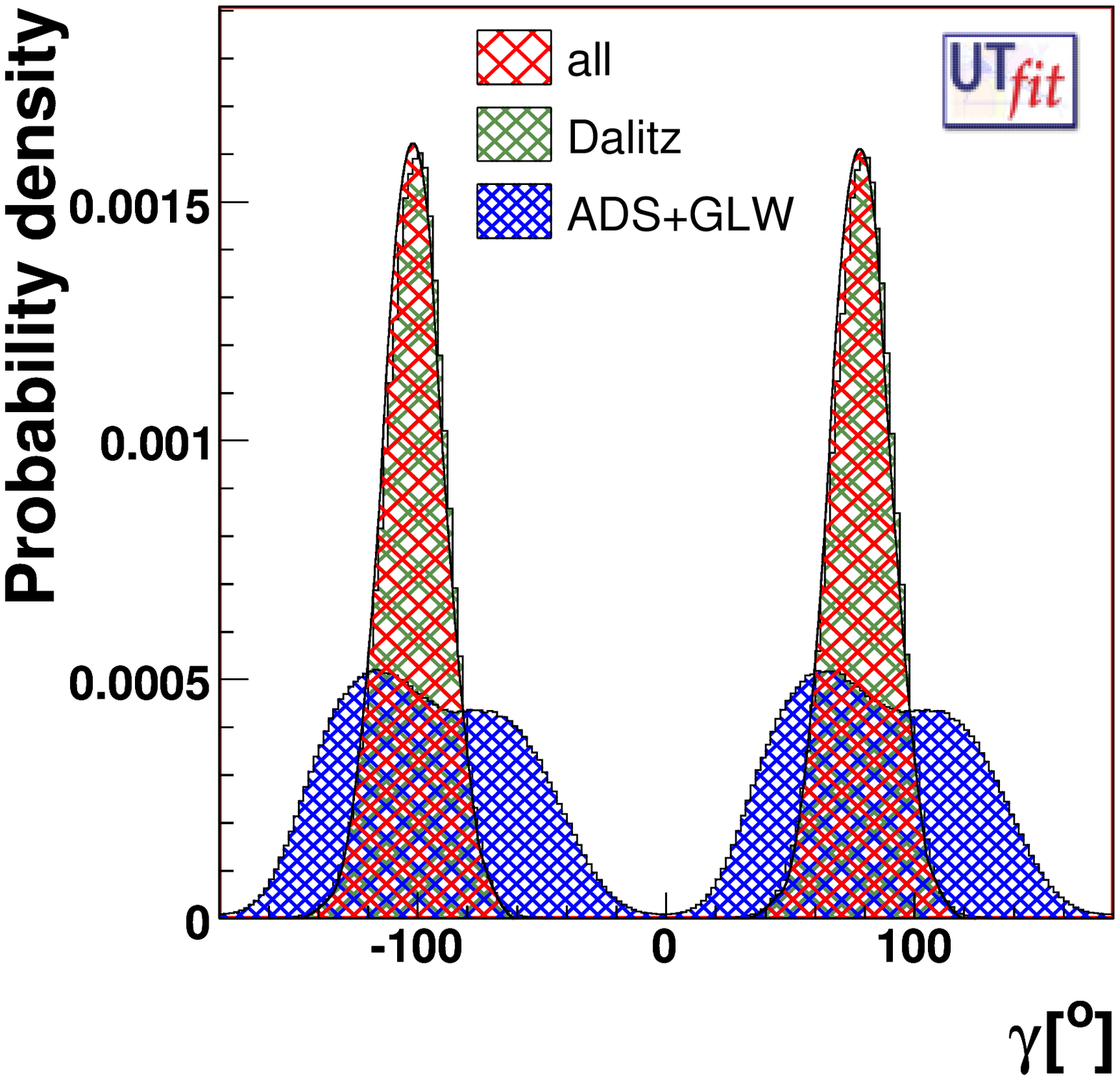}
\includegraphics[width=6cm,height=5.9cm]{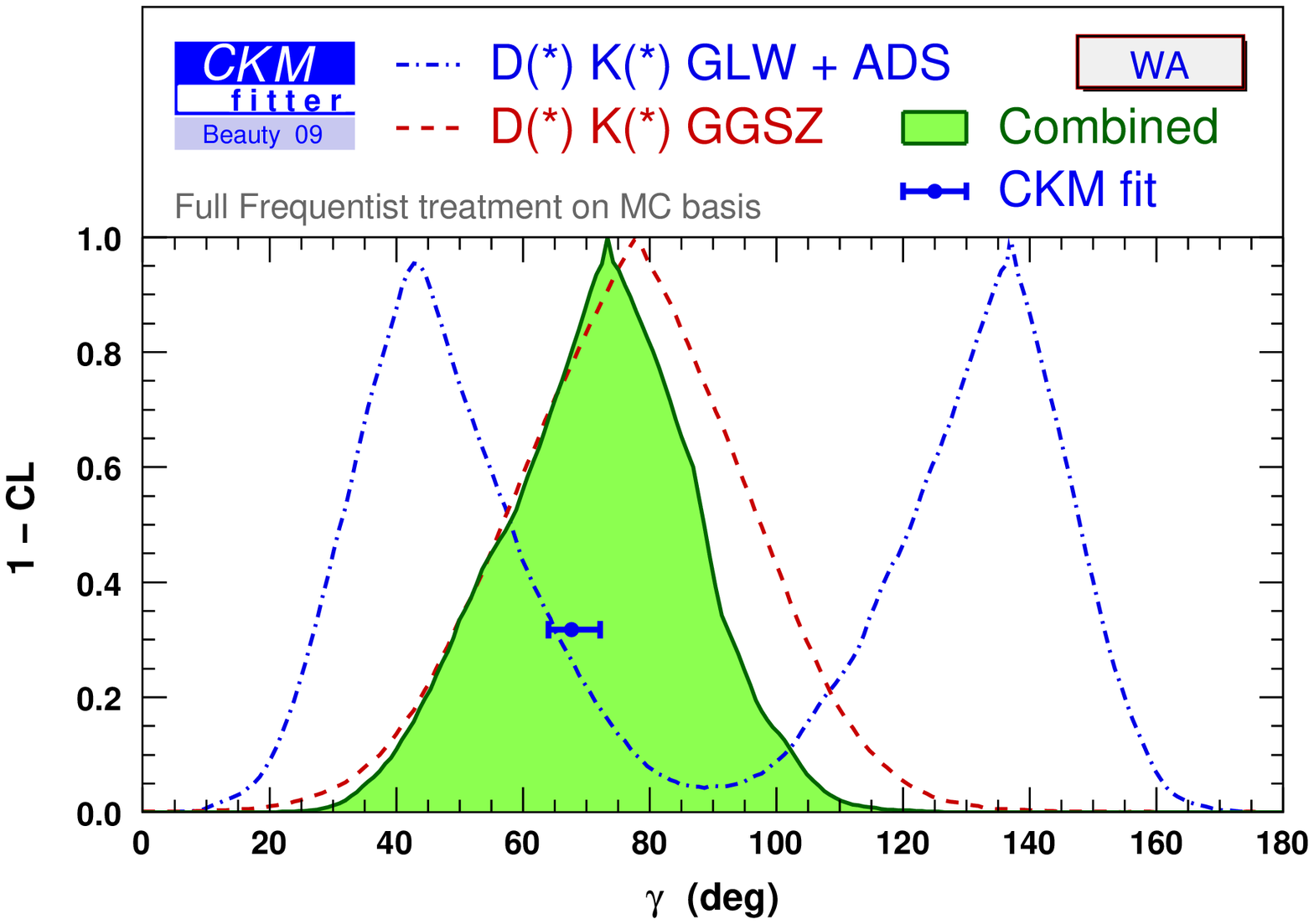}
\end{center}
\caption{Left: probability density function of $\gamma$ from the combination of BaBar and Belle measurements of $B^\pm\to D^{(*)0}K^{(*)\pm}$ decays as obtained by UTfit~\cite{utfit}. Right: 1-confidence-level as derived by CKMfitter~\cite{ckmfitter}.} 
\label{fig:combination}
\end{figure}

Using a Bayesian statistical procedure and combining BaBar and Belle results of GLW, ADS and Dalitz methods, the UTfit collaboration finds $\gamma=(78\pm 12)^\circ$~\cite{utfit}\footnote{The average is not updated with the most recent BaBar's $B\to D^{(*)0}K^{(*)}$ ADS and Belle's Dalitz measurements.}, while CKMfitter using a frequentist approach quotes $\gamma= (73^{+22}_{-25})^\circ$~\cite{ckmfitter}. The results are reported in Fig.~\ref{fig:combination}. To reduce the error to a few degrees we have to wait for LHCb ($\sigma_\gamma\approx 2-3^\circ$ with 10~fb$^{-1}$~\cite{lhcb}) or a Super Flavor factory ($\sigma_\gamma\approx 1^\circ$ with 75ab$^{-1}$~\cite{cdr_superb}). 
 
\end{document}